\documentclass[fleqn,10pt]{wlscirep}

\usepackage{amssymb}
\usepackage{amsfonts}
\usepackage{amsmath}

\usepackage{amsbsy}
\usepackage{amsthm}
\usepackage{bbm}

\usepackage{color}

\usepackage{epsfig}
\usepackage{graphicx}
\usepackage{mathrsfs}

\newcommand{\beq}{\begin{eqnarray}}
\newcommand{\eeq}{\end{eqnarray}}

\newcommand{\tr}{\text{tr}}

\newcommand{\be}{\begin{equation}}
\newcommand{\ee}{\end{equation}}
\newcommand{\bea}{\begin{eqnarray}}
\newcommand{\eea}{\end{eqnarray}}
\newcommand{\bes}{\begin{equation*}}
\newcommand{\ees}{\end{equation*}}
\newcommand{\beas}{\begin{eqnarray*}}
\newcommand{\eeas}{\end{eqnarray*}}

\def\v{\vec{v}}

\def\<{\langle}
\def\>{\rangle}

\def\tr{\mathrm{tr}}

\newtheorem{Proposition}{Proposition}

\theoremstyle{definition}
\newtheorem{ex}{Example}

\newtheorem{defn}{Definition}
\newtheorem{remark}{Remark}

\newtheorem*{lipschitzLem*}{Lemma \ref{lipschitz}}
\newtheorem*{lipschitzCubeLem*}{Lemma \ref{lipschitzCube}}
\newtheorem*{pgmNearlyOptimalThm*}{Theorem \ref{pgmNearlyOptimal}}

\newcommand{\cc}{\mathbb{C}}

\newcommand{\bei}{\begin{itemize}}
\newcommand{\eei}{\end{itemize}}

\def\oper{{\mathchoice{\rm 1\mskip-4mu l}{\rm 1\mskip-4mu l}{\rm 1\mskip-4.5mu l}{\rm 1\mskip-5mu l}}}

\begin{document}

\title{A mirrored pair of optimal non-decomposable entanglement witnesses for two qudits does exist}

\author[1]{Dariusz Chru{\'s}ci{\'n}ski}
\author[2]{Anindita Bera}
\author[3]{Joonwoo Bae}
\author[4,*]{Beatrix C. Hiesmayr}

\affil[1]{Institute of Physics, Faculty of Physics, Astronomy and Informatics,
Nicolaus Copernicus University, Grudzi\c{a}dzka 5/7, 87--100 Toru{\'n}, Poland}
\affil[2]{Department of Mathematics, Birla Institute of Technology Mesra, Jharkhand 835215, India}
\affil[3]{School of Electrical Engineering, Korea Advanced Institute of Science and Technology (KAIST),
291 Daehak-ro, Yuseong-gu, Daejeon 34141, Republic of Korea}
\affil[4]{University of Vienna, Faculty of Physics, Boltzmanngasse 5, 1090 Vienna, Austria}

\affil[*]{Beatrix.Hiesmayr@univie.ac.at}

\begin{abstract}
Two approaches can be utilized to handle the separability problem, finding out whether a given bipartite qudit state is separable or not: a direct procedure on the state space or the effective tool of entanglement witnesses (EWs). This contribution studies the structure of EWs. Exploiting the very concept of mirrored EWs, increasing the detection power, we show, in contrast to the conjecture in a recent paper (Sci. Rep. {\bf 13}, 10733 (2023)), there exist pairs of optimal EWs, which are both non-decomposible, i.e. can detect bound/PPT-entangled states in an optimal way. Since we show that the structure also extends to higher dimensions, our results reveal a further structure of entanglement witnesses.
\end{abstract}

\maketitle

\section{Introduction}

Entanglement witnesses (EWs) are both theoretical and experimental tools to detect entangled states \cite{Terhal2002,HHHH,EW2,TOPICAL,KYE,ani18}. An EW is represented by Hermitian operator $W$ in $\mathcal{H}_A \otimes \mathcal{H}_B$ such that $W$ is not positive definite (possesses at least one negative eigenvalue) but $\< \psi_A \otimes \psi_B|W|\psi_A \otimes \psi_B\> \geq 0$ for all product vectors (such operators are often called to be block-positive).  It immediately implies that  $\tr[W\sigma_{\mathrm{sep}}] \geq 0$ for all separable states $\sigma_{\mathrm{sep}}$ and hence for any entangled state $\rho$
there exists an EWs $W$ such that $\tr[W\rho]<0$ \cite{Terhal2002,HHHH}.

There exists a parallel detection scheme based on the concept of positive maps \cite{HHHH}.
Due to the well-known Choi-Jamio{\l}kowski isomorphism \cite{Pillis,Jam,MDChoi} (cf. \cite{Karol} for a detailed exposition), there is an one-to-one correspondence between the block-positive operators in $\mathcal{H}_A \otimes \mathcal{H}_B$ and positive linear maps $\Phi : \mathcal{B}(\mathcal{H}_A) \to \mathcal{B}(\mathcal{H}_B)$, where $\mathcal{B}(\mathcal{H})$ denotes bounded linear operators acting on $\mathcal{H}$. Recall that $\Phi$ is a positive map whenever for any $X \geq 0$ one has $\Phi(X) \geq 0$. Moreover, if $\mathcal{I}_n \otimes \Phi$ is also positive for all $n=1,2,\ldots$, then $\Phi$ is called completely positive \cite{Paulsen,Stormer} (here  $\mathcal{I}_n$ denotes an identity map acting on a space of $n\times n$ complex matrices). Completely positive maps are fully characterized by the celebrated Kraus representation $\Phi(X) = \sum_\alpha K_\alpha X K_\alpha^\dagger$
with suitable Kraus operators $K_\alpha$. Entanglement witnesses correspond to positive but not completely positive maps \cite{Terhal2002,HHHH,EW2,TOPICAL, ani-giovanni}. Unfortunately, in spite of the considerable effort, the structure of positive maps -- and hence also entanglement witnesses -- is still not fully understood. This, in turn, implies that the full classification of quantum states of composite systems is still missing.

An EW is called decomposable if $W=A + B^\Gamma$, with $A,B \geq 0$ and $\Gamma$ stands for the partial transposition. Otherwise, it is called non-decomposable. Decomposable witnesses cannot detect PPT entangled states, i.e. if $W = A + B^\Gamma$, then ${\rm Tr}(W\sigma_{\rm PPT})\geq 0 $ for all PPT states $\sigma_{\rm PPT}$. Hence, to classify bound entangled states, the construction of large classes of non-decomposable witnesses is of great importance. A recent review on the current knowledge about bound entangled or PPT entangled states can be found in Ref.~\cite{ReviewBoundHiesmayr}.

Among EWs, optimal witnesses are of particular importance \cite{Lew}. An EW $W$ is called optimal if, for any $P \geq 0$, the following operator $W -  P$ is no longer an EW. It means that one cannot improve $W$ by subtracting a positive operator.
Entanglement witnesses, being Hermitian operators, represent physical observables and, hence, in principle, can be implemented in the laboratory. Moreover, any EW can be  factored into local observables
\bea
W = \sum_{i}  A_i\otimes B_i ,
\label{eq:lo}
\eea
with  local observables $A_i$ and $B_i$. A collection of expectation values of local observables $\tr[A_i\otimes B_j \rho]$ can decide whether a state $\rho$ is entangled.

\begin{figure}
  \includegraphics[width=0.8\textwidth]{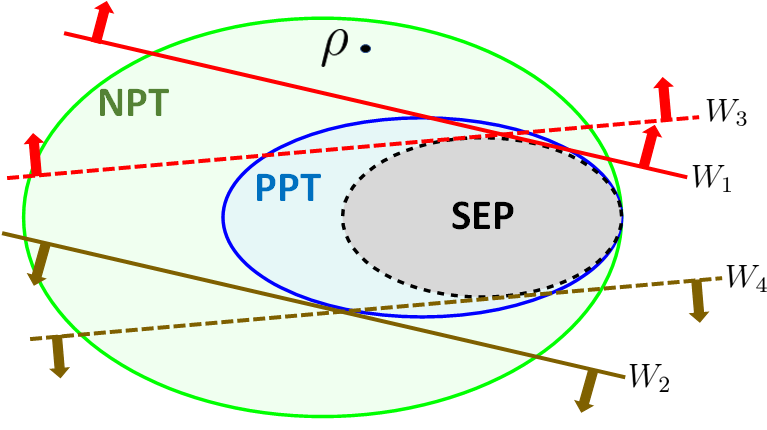}
  \caption{This graphic displays the set of bipartite states (green) together with convex subsets of PPT states (blue) and separable states (grey). States which are PPT but not separable are bound entangled.
Entanglement witnesses $W_i$ are represented by straight lines. In Ref.~\cite{Anindita-23} the conjecture was put forward based on the study of several known examples, that for a pair of mirrored EWs, $\{W_1,W_2\}$
if $W_1$ is optimal and non-decomposable then $W_2$ is decomposable (or it's not a witness at all).
This paper shows that surprisingly pairs of mirrored optimal EWs exist, here $\{W_3,W_4\}$ (dotted lines), which both detect PPT-entangled states.}
    \label{Fig-J}
\end{figure}

Recently, in Ref.~\cite{MEW} the framework of mirrored EWs was introduced. Given an entanglement witness $W$ one defines a mirrored operator $W_{\rm M}$ by
\begin{equation}   \label{M!}
    W_{\rm M} = \mu\; \oper_A \otimes \oper_B - W,
\end{equation}
with the smallest $\mu>0$ such that $W_{\rm M}$ is block-positive, i.e. $\< \psi\otimes \phi|W_{\rm M}|\psi \otimes \phi\> \geq 0$. Now, if the maximal eigenvalue $\lambda_{\rm max}$ of $W$ satisfies $\lambda_{\rm max} > \mu$, then $W_{\rm M}$ is an EW and therefore this construction gives rise to a pair $(W,W_{\rm M})$ of mirrored EWs~\cite{MEW},  which can double up the capability of detecting entangled states. Interestingly, the construction of a mirrored operator $W_{\rm M}$ is closely related to the well-known structural physical approximation \cite{SPA-01,SPA-02,SPA-1,SPA-2,SPA-3,SPA-4}. In a recent paper~\cite{Anindita-23} we constructed mirrored operators for several classes of well-known EWs both decomposable and non-decomposable. In particular, it was shown that a mirrored operator to any extremal  decomposable EW is always a positive operator (cf. the next section for the definition of optimal and extremal EWs). We posed the conjecture that the mirrored operator obtained from an optimal EW is either a positive operator or a decomposable EW.
 The conjecture~\cite{Anindita-23} was motivated by the simple observation that there exists a trade-off relation between $W$ and $W_{\rm M}$ displayed by eq. (\ref{M!}). In particular, it says that there does not exist a mirrored pair of non-decomposable EWs $(W,W_{\rm M})$ such that at least one of them is optimal. The assumption about optimality turned out to be critical since we provided an example of a mirrored pair of non-decomposable EWs, but none of them was optimal. In  Fig.~\ref{Fig-J} we have visualized our findings, the graphic schematically displays the convex subsets of separable and PPT states and two mirrored pairs of EWs represented by parallel lines. The (dotted) pair of EWs $\{W_3,W_4\}$ was conjectured not to exist.

In this paper, we report a counterexample to the above conjecture, providing a mirrored pair of two optimal non-decomposable EWs for two qutrits. Interestingly, both EWs have maximal numbers of negative eigenvalues and the corresponding eigenvectors span a completely entangled subspace. This example further reveals the intricate structure of EWs. Possible examples of mirrored non-decomposable optimal EWs in higher dimensions are also discussed.

The paper is organized as follows. In Sec.~\ref{sec:MEW}, we briefly review the concept of optimal EWs. Section \ref{Results} presents the construction of a mirrored pair of non-decomposable witnesses. Our main results state that they are both optimal (even non-decomposable optimal \cite{Lew,Lew-2}). Section \ref{Methods} contains technical details of the proof. Section~\ref{SEC-C} discusses generalization for higher dimensional systems. We finally conclude with open questions in Sec.~\ref{SEC-SUM}.

\section{  Optimal entanglement witnesses }\label{sec:MEW}

In this section, we briefly discuss the very concept of optimal EWs \cite{Lew,Lew-2}. Let $\mathcal{D}_{W}$ denote the set of states detected by an EW $W$, i.e. $\rho \in \mathcal{D}_W$ if  $\tr(W\rho) < 0$. Now, given two EWs $W_1$ and $W_2$ one says that $W_1$ is finer than $W_{2}$ if $ D_{W_{2}} \subseteq D_{W_{1}}$ \cite{Lew}. An EW $W$ is optimal if there is no EW finer than $W$.

It is shown \cite{Lew,Lew-2} that if $W$ is optimal, then $W-P$ is no longer an EW, where $P$ is an arbitrary positive operator.
This means that one cannot improve $W$ (i.e. make it finer)  by subtracting $P \geq 0$. Note that optimality does not protect to subtract a block-positive operator. Finally, $W$ is extremal if and only if $W-B$ is no longer an EW, where $B$ is an arbitrary block-positive operator such that $B \neq \lambda W$.
Clearly, any extremal EW is optimal. However, the converse need not be true. A key example of an optimal decomposable EW is the swap (flip) operator in $\mathbb{C}^d \otimes \mathbb{C}^d$

\begin{equation}
    \mathbb{F} = d\; {P^{+\Gamma}_d} = \sum_{k,l=1}^d |k\>\<\ell| \otimes |\ell\>\<k| ,
\end{equation}
where $P^+_d$ denotes the canonical maximally entangled state in $\mathbb{C}^d \otimes \mathbb{C}^d$.
However, an EW corresponding to so-called reduction map $ R_d(X) = \oper_d\; \tr X - X $ defined by
\begin{equation} \label{Rn}
   W= \oper_d \otimes \oper_d - dP ^+_d ,
\end{equation}
is optimal for all $d\geq 2$ but extremal only for $d=2$ \cite{HHHH,TOPICAL}.

In general, given $W$, it is very hard to check whether it is optimal. There exists, however, an operational sufficient condition for optimality \cite{Lew}. Denote by $P_W$ a set of product vectors $|\psi \otimes \phi\rangle$ such that
\begin{equation}
    \< \psi \otimes \phi |W| \psi \otimes \phi \> = 0 .
\end{equation}
One has the following \cite{Lew}
\begin{Proposition} \label{PRO-1} If ${\rm span}\, P_W = \mathcal{H}_A \otimes \mathcal{H}_B$, then $W$ is optimal.
\end{Proposition}
In this case, i.e. when ${\rm span}_{\mathbb{C}}P_W = \mathcal{H}_A \otimes \mathcal{H}_B$, one says that $W$ has the spanning property. It should be stressed, however, that there exist optimal EWs without spanning property (cf. recent discussion in Ref.~\cite{LAA}).  A key result concerning the
 structure of optimal positive maps was derived in \cite{maciej00,philipPRA,aniarxiv}.

Consider a decomposable EW $W = A + B^\Gamma$ in $\mathbb{C}^n \otimes \mathbb{C}^m$. Recall that $W$ is optimal if $W = B^\Gamma$ and $B$ is supported on a completely entangled subspace (CES) \cite{maciej00}. A linear subspace $\Sigma \subset \mathbb{C}^n \otimes \mathbb{C}^m$ defines a CES if it does not contain a product vector. It is well known that a maximal dimension of any CES in  $\mathbb{C}^n \otimes \mathbb{C}^m$ is $(n-1)(m-1)$ \cite{CES1,CES1a,CES2} The simplest example of CES is a 1-dimensional subspace spanned by an arbitrary entangled vector $|\Psi\rangle \in \mathbb{C}^n \otimes \mathbb{C}^m$. The corresponding entanglement witness $|\Psi\>\<\Psi|^\Gamma$ is extremal \cite{TOPICAL}. It is, therefore, clear that any decomposable EW is a convex combination of extremal witnesses.

For non-decomposable EWs, the situation is much more complicated \cite{jpa22}. Recall that a bipartite state is called a PPT state (Positive Partial Transpose) if $\rho^\Gamma \geq 0$, i.e. both $\rho$ and $\rho^\Gamma$ are legitimate quantum states. Now, $W$ is a non-decomposable EW if and only if it detects some PPT-entangled states. Let $\mathcal{D}^{\rm PPT}_W$ be a set of PPT states detected by $W$.
Now,  $W_1$ is non-decomposable--finer than $W_2$ if  $\mathcal{D}^{\rm PPT}_{W_2} \subseteq \mathcal{D}^{\rm PPT}_{W_1}$.
An EW $W$ is non-decomposable--optimal if there is no non-decomposable--finer EW than $W$. Actually, if an EW $W$ is non-decomposable--optimal, then $W-D$ for a PPT operator $D$ is no longer an EW.
It means that one cannot improve $W$ (i.e. make it finer)  by subtracting a PPT operator $P$. Interestingly, it has been proven~\cite{Lew}

\begin{Proposition}
$W$ is non-decomposable--optimal if and only if both $W$ and $W^\Gamma$ are optimal.
\end{Proposition}
Now, $W$ has a bi-spanning property if there exists a set of product vectors $\psi_k \otimes \phi_k$ such that $\<\psi_k \otimes \phi_k|W|\psi_k \otimes \phi_k\>=0$ together with

$$  {\rm span} \{\psi_k \otimes \phi_k\} = {\rm span} \{ \psi_k \otimes \phi_k^*\} = \mathcal{H}_A \otimes \mathcal{H}_B  . $$
In analogy to Proposition \ref{PRO-1} one proves \cite{Lew}
\begin{Proposition}
    If $W$ has a bi-spanning property, then it is non-decomposable optimal.
\end{Proposition}
EWs displaying a bi-spanning property were analyzed in Refs.~\cite{Ha1,Ha2,Ha3}.

\section{Results}  \label{Results}

Our main result is based on the construction of mirrored entanglement witnesses based on mutually unbiased bases (for a recent review, see~\cite{ReviewMUBs}). Two orthonormal bases $\{\psi_k\}$ and $\{\phi_\ell\}$ in $\mathbb{C}^d$ are mutually unbiased (MUB) if

\begin{equation}
    |\< \psi_k|\phi_\ell\>|^2 = \frac 1 d \ , \ \ \ k,\ell=1,2,\ldots, d .
\end{equation}
It is well known \cite{Wootters} that in $\mathbb{C}^d$ there are at most `$d+1$'  MUBs. In particular, if $d$ is a power of prime, then it is known how to construct the maximal set of `$d+1$' MUBs. For $d=3$ one has four MUBs $ \mathcal{B}_1,\ldots, \mathcal{B}_4$ defined as follows:   $ \mathcal{B}_1 =  \{ \psi^{(1)}_1=|1\>,\psi^{(1)}_2=|2\>,\psi^{(1)}_3=|3\>\}$, where $|1\>,|2\>,|3\>$ defines a computational basis in $\mathbb{C}^3$, together with
$\mathcal{B}_2$, $\mathcal{B}_3$ and $\mathcal{B}_4$:

\begin{eqnarray*}
  && \mathcal{B}_2 = \left\{ \frac{|1\> + |2\> +|3\>}{\sqrt{3}},  \frac{|1\> + \omega^* |2\> + \omega|3\>}{\sqrt{3}} ,  \frac{|1\> + \omega |2\> + \omega^* |3\>}{\sqrt{3}} \right\} , \\
   && \mathcal{B}_3 = \left\{ \frac{|1\> + |2\> + \omega^* |3\>}{\sqrt{3}},  \frac{|1\> + \omega |2\> + \omega|3\>}{\sqrt{3}} ,  \frac{|1\> + \omega^* |2\> + |3\>}{\sqrt{3}} \right\} , \\
  && \mathcal{B}_4 =  \left\{ \frac{|1\> + |2\> + \omega |3\>}{\sqrt{3}},  \frac{|1\> + \omega^* |2\> + \omega^* |3\>}{\sqrt{3}} ,  \frac{|1\> + \omega |2\> + |3\>}{\sqrt{3}} \right\} ,
\end{eqnarray*}
with $\omega = e^{2 \pi i/3}$. With the projection operators $P^{(\alpha)}_k := |\psi^{(\alpha)}_k\> \< \psi^{(\alpha)}_k|$ we can define the following four quantum channels
\begin{equation}
    \Phi_\alpha(\rho) := \sum_{k=1}^3 P^{(\alpha)}_k \rho P^{(\alpha)}_k \;.
\end{equation}
In particular,
\begin{equation}
    \Phi_1(\rho) = \sum_{k=1}^3 |k\>\<k| \rho |k\>\<k|
\end{equation}
describes complete decoherence of $\rho$ w.r.t. the computational basis. Similarly, $ \Phi_\alpha$ describes complete decoherence of $\rho$ w.r.t. $\{ |\psi^{(\alpha)}_k\>\} $. Finally, let
\begin{equation}
    \Phi_0(\rho) := \frac 13\;  \oper\, {\rm Tr}\rho ,
\end{equation}
denote a completely depolarizing channel. Now, let us split a set $\{1,2,3,4\}$ into two disjoint 2-element sets $\Gamma$ and its complement $\Gamma_c$, i.e. $\Gamma \cup \Gamma_c = \{1,2,3,4\}$ and define the map
\begin{equation}  \label{Phi!}
    \Phi_\Gamma := 2\, \Phi_0 +  \sum_{\alpha \in \Gamma_c}  \Phi_\alpha  - \sum_{\beta \in \Gamma} \Phi_\beta .
\end{equation}
One proves \cite{MUB-Filip,Kasia} that $\Phi_\Gamma$ defines a positive  map, or, equivalently, the corresponding Choi matrix
\begin{equation}
    W_\Gamma = \sum_{k,\ell=1}^3 |k\>\<\ell| \otimes \Phi_\Gamma(|k\>\<\ell|) ,
\end{equation}
defines an entanglement witness.

\begin{Proposition} $W_\Gamma$ and $W_{\Gamma_c}$ define a mirrored pair of non-decomposable EWs.
\end{Proposition}
Indeed, one has
\begin{equation}
     \Phi_\Gamma +  \Phi_{\Gamma_c} = 4\; \Phi_0 ,
\end{equation}
or, equivalently,
\begin{equation}
     W_\Gamma +  W_{\Gamma_c} = 4\; \oper \otimes \oper .
\end{equation}
In particular for $\Gamma = \{1,2\}$ one finds

\begin{equation}  \label{W&W}
{W}_\Gamma =
\left[\begin{array}{c c c|c c c|c c c}
\cdot & \cdot & \cdot & \cdot & 1 & \cdot & \cdot & \cdot & 1 \\
\cdot & 3 & \cdot & \cdot & \cdot & -2 & -2 & \cdot & \cdot \\
\cdot & \cdot & 3 & -2 & \cdot & \cdot & \cdot & -2 & \cdot \\
\hline
\cdot & \cdot & -2 & 3 & \cdot & \cdot & \cdot & -2 & \cdot \\
1 & \cdot & \cdot & \cdot & \cdot & \cdot & \cdot & \cdot & 1 \\
\cdot & -2 & \cdot & \cdot & \cdot & 3 & -2 & \cdot & \cdot \\
\hline
\cdot & -2 & \cdot & \cdot & \cdot & -2 & 3 & \cdot & \cdot \\
\cdot & \cdot & -2 & -2 & \cdot & \cdot & \cdot & 3 & \cdot \\
1 & \cdot & \cdot & \cdot & 1 & \cdot & \cdot & \cdot & \cdot
\end{array}\right],
\qquad
{W}_{\Gamma_c} =
\left[\begin{array}{c c c|c c c|c c c}
4 & \cdot & \cdot & \cdot & -1 & \cdot & \cdot & \cdot & -1 \\
\cdot & 1 & \cdot & \cdot & \cdot & 2 & 2 & \cdot & \cdot \\
\cdot & \cdot & 1 & 2 & \cdot & \cdot & \cdot & 2 & \cdot \\
\hline
\cdot & \cdot & 2 & 1 & \cdot & \cdot & \cdot & 2 & \cdot \\
-1 & \cdot & \cdot & \cdot & 4 & \cdot & \cdot & \cdot & -1 \\
\cdot & 2 & \cdot & \cdot & \cdot & 1 & 2 & \cdot & \cdot \\
\hline
\cdot & 2 & \cdot & \cdot & \cdot & 2 & 1 & \cdot & \cdot \\
\cdot & \cdot & 2 & 2 & \cdot & \cdot & \cdot & 1 & \cdot \\
-1 & \cdot & \cdot & \cdot & -1 & \cdot & \cdot & \cdot & 4
\end{array}\right],
\end{equation}
where to make the matrices more transparent, we replaced all zeros with dots. The corresponding pair of PPT entangled states detected by $W_\Gamma$ and $W_{\Gamma_c}$ reads

\begin{equation}
\rho_{\Gamma} = \frac{1}{15}
\left[\begin{array}{c c c|c c c|c c c}
3 & \cdot & \cdot & \cdot & \cdot & \cdot & \cdot & \cdot & \cdot \\
\cdot & 1 & \cdot & \cdot & \cdot & 1 & 1 & \cdot & \cdot \\
\cdot & \cdot & 1 & 1 & \cdot & \cdot & \cdot & 1 & \cdot \\
\hline
\cdot & \cdot & 1 & 1 & \cdot & \cdot & \cdot & 1 & \cdot \\
\cdot & \cdot & \cdot & \cdot & 3 & \cdot & \cdot & \cdot & \cdot \\
\cdot & 1 & \cdot & \cdot & \cdot & 1 & 1 & \cdot & \cdot \\
\hline
\cdot & 1 & \cdot & \cdot & \cdot & 1 & 1 & \cdot & \cdot \\
\cdot & \cdot & 1 & 1 & \cdot & \cdot & \cdot & 1 & \cdot \\
\cdot & \cdot & \cdot & \cdot & \cdot & \cdot & \cdot & \cdot & 3
\end{array}\right]
\quad
\rho_{\Gamma_c} = \frac{1}{15} \left[
\begin{array}{ccc|ccc|ccc}
 1 & \cdot & \cdot & \cdot & 1 & \cdot & \cdot & \cdot & 1 \\
 \cdot & 2 & \cdot & \cdot & \cdot & -1 & -1 & \cdot & \cdot \\
 \cdot & \cdot & 2 & -1 & \cdot & \cdot & \cdot & -1 & \cdot \\\hline
 \cdot & \cdot & -1 & 2 & \cdot & \cdot & \cdot & -1 & \cdot \\
 1 & \cdot & \cdot & \cdot & 1 & \cdot & \cdot & \cdot & 1 \\
 \cdot & -1 & \cdot & \cdot & \cdot & 2 & -1 & \cdot & \cdot \\\hline
 \cdot & -1 & \cdot & \cdot & \cdot & -1 & 2 & \cdot & \cdot \\
 \cdot & \cdot & -1 & -1 & \cdot & \cdot & \cdot & 2& \cdot \\
 1 & \cdot & \cdot & \cdot & 1 & \cdot & \cdot & \cdot & 1 \\
\end{array}
\right].
\end{equation}
Straightforwardly, one easily finds
\begin{equation}
    {\rm Tr}\left(W_\Gamma \rho_\Gamma\right) \, =\, {\rm Tr}\left(W_{\Gamma_c} \rho_{\Gamma_c} \right) = - \frac 25 .
\end{equation}

All such EWs $W_\Gamma$ with 2-element set $\Gamma$ have the following circulant structure
\begin{equation}
W_\Gamma =\left(
\begin{array}{ccc|ccc|ccc}
 a & \cdot & \cdot & \cdot & x & \cdot & \cdot & \cdot & x \\
 \cdot & b & \cdot & \cdot & \cdot & z & z^* & \cdot & \cdot \\
 \cdot & \cdot & b & z^* & \cdot & \cdot & \cdot & z & \cdot \\\hline
 \cdot & \cdot & z & b & \cdot & \cdot & \cdot & z^* & \cdot \\
 x & \cdot & \cdot & \cdot & a & \cdot & \cdot & \cdot & x \\
 \cdot & z^* & \cdot & \cdot & \cdot & b & z & \cdot & \cdot \\\hline
 \cdot & z & \cdot & \cdot & \cdot & z^* & b & \cdot & \cdot \\
 \cdot & \cdot & z^* & z & \cdot & \cdot & \cdot & b & \cdot \\
 x & \cdot & \cdot & \cdot & x & \cdot & \cdot & \cdot & a \\
\end{array}
\right),
\end{equation}
where the parameters $\{a,b,x,z\}$ depend upon $\Gamma$ as presented in Table \ref{T1}

\begin{table}[h!]
\centering
\begin{tabular}{|c|c|c|c|c|}
\hline
$\Gamma$ &  $a$ & $b$ & $x$ & $z$ \\ \hline
$\ \{1,2\}\ $     &\  0 \    & \ 3 \     &\ 1\      & $-2$       \\ \hline
$\ \{1,3\}\ $      &\  0  \    & \ 3\      &\ 1 \      & $1 - i \sqrt{3}$       \\ \hline
$\ \{1,4\}\ $       &\  0 \     & \ 3\      &\ 1\      & $\ 1 + i \sqrt{3}\ $       \\ \hline
$\ \{3,4\}\ $      &\  4 \    &\ 1\      & $-1$       & $2$      \\ \hline
$\ \{2,4\}\ $      &\  4 \     &\ 1\      & $-1$      & $-1 + i \sqrt{3} $      \\ \hline
$\ \{2,3\}\ $      &\ 4  \    &\ 1\      & $\ -1\ $     &  $\ -1 - i \sqrt{3} \ \ $      \\ \hline
\end{tabular}
\caption{Parameters $\{a,b,x,z\}$ for different 2-element sets $\Gamma$}
\label{T1}
\end{table}

It is well known \cite{TOPICAL} that any EW in $\mathbb{C}^n \otimes \mathbb{C}^m$ may have at most $(n-1)\times (m-1)$ strictly negative eigenvalues. Let us observe that witness $W_\Gamma$ saturates this bound, one finds for its spectrum
$$ \{5, 5, 5, 5, 2, -1, -1, -1, -1\} ,$$
that is, it has exactly $(d-1)^2=4$ negative eigenvalues. It should be stressed that the spectrum does not depend upon the choice of the 2-element set $\Gamma$. All EWs $W_\Gamma$ are isospectral. Another interesting observation is that four eigenvectors corresponding to the eigenvalue `$-1$' span a completely entangled subspace (CES) in $\mathbb{C}^3 \otimes \mathbb{C}^3$ \cite{CES1,CES1a,CES2}. Recall that a linear subspace $S \subset  \mathbb{C}^n \otimes \mathbb{C}^m$ defines a CES if there is no product vector in $S$. Again, the maximal dimension of CES is $(n-1)\times (m-1)$.

\begin{Proposition} Entanglement witness $(W_\Gamma,W_{\Gamma_c})$ are non-decomposable optimal.
\end{Proposition}
It is sufficient to show non-decomposable optimality for e.g. $\Gamma=\{1,2\}$. Following Ref.~\cite{Lew} we show that there exists a set of product vectors $|\alpha_k \otimes \beta_k\>$ $(k=1,2,\ldots,9)$ such that
\begin{equation}\label{W=0}
    \< \alpha_k \otimes \beta_k|W_\Gamma|\alpha_k \otimes \beta_k\> = 0 ,
\end{equation}
and both sets $\{  |\alpha_k \otimes \beta_k\> \}_{k=1}^9 $ and $\{  |\alpha_k \otimes \beta_k^*\>\}_{k=1}^9 $ span $\mathbb{C}^3 \otimes \mathbb{C}^3$. The following set of product vectors does the job:

\begin{equation}   \label{VECTORS}
\begin{array}{ll} \vspace{.1cm}
     |\alpha_1\> = |1\> \ , \ & |\beta_1\> = |1\>  \ ,\\ \vspace{.1cm}
 |\alpha_2\> = |1\> + \xi (|2\> + |3\> )  \ , \ & |\beta_2\> = |1\> + \xi (|2\> + |3\> )  \ ,\\ \vspace{.1cm}
    |\alpha_3\> = |1\> + \xi ( \omega |2\> + \omega^* |3\> )  \ , \ & |\beta_3\> = |1\> + \xi (\omega^* |2\> + \omega |3\> )  \ ,\\ \vspace{.1cm}
     |\alpha_4\> = |2\> \ , \ & |\beta_4\> = |2\>  \ ,\\ \vspace{.1cm}
     |\alpha_5\> = |2\> + \xi (|1\> + |3\> )  \ , \ & |\beta_5\> = |2\> + \xi (|1\> + |3\> )  \ ,\\ \vspace{.1cm}
     |\alpha_6\> = |2\> + \xi ( \omega |1\> + \omega^* |3\> )  \ , \ & |\beta_6\> = |2\> + \xi (\omega^* |1\> + \omega |3\> )  \ , \\ \vspace{.1cm}
     |\alpha_7\> = |3\> \ , \ & |\beta_7\> = |3\>  \ ,\\ \vspace{.1cm}
     |\alpha_8\> = |3\> + \xi (|1\> + |2\> )  \ , \ & |\beta_8\> = |3\> + \xi (|1\> + |2\> )  \ ,\\ \vspace{.1cm}
     |\alpha_9\> = |3\> + \xi ( \omega |1\> + \omega^* |2\> )  \ , \ & |\beta_9\> = |3\> + \xi (\omega^* |1\> + \omega |2\> )  \ ,
\end{array}
\end{equation}
where $\xi = \frac{1}{\sqrt{2}} e^{i\pi/4}$. One easily checks that indeed (\ref{W=0}) holds for any $|\alpha_k \otimes \beta_k\>$ with $k=1,2,\ldots,9$. In the next section, we analyze the above set in more detail and prove the bi-spanning property of $W_\Gamma$.

Note that $W_\Gamma$ and $W_{\Gamma_c}$ are iso-spectral and hence they are locally unitarily equivalent. One finds $W_{\Gamma_c} =U \otimes U^*\; W_{\Gamma}\; U^\dagger\otimes U^{*\dagger}$, where
\begin{eqnarray}
    U=\frac{1}{\sqrt{3}}\left(
\begin{array}{ccc}
 1 & 1 & \omega\\
 1 & \omega & 1 \\
 \omega^* & \omega & \omega \end{array}
\right) ,
\end{eqnarray}
and hence the following vectors
\begin{eqnarray}
    |\tilde{\alpha}_k \otimes \tilde{\beta_k}\> := |U\alpha_k \otimes U^*\beta_k\> ,
\end{eqnarray}
define a bi-spanning family for $W_{\Gamma_c}$, that is,
\begin{itemize}
    \item $\< \tilde{\alpha}_k \otimes \tilde{\beta_k} |W_{\Gamma_c}| \tilde{\alpha}_k \otimes \tilde{\beta_k}\> = 0$ for $k=1,2,\ldots,9$
    \item vectors $|\tilde{\alpha}_k \otimes \tilde{\beta_k}\>$ span $\mathbb{C}^3 \otimes \mathbb{C}^3$, and  vectors $|\tilde{\alpha}_k \otimes \tilde{\beta_k}^*\>$ also span $\mathbb{C}^3 \otimes \mathbb{C}^3$.
\end{itemize}

\section{Methods}   \label{Methods}

It is clear that since $(W_\Gamma)_{kk,kk}=0$ one has
\begin{equation}  \label{I}
    \< k \otimes k|W_\Gamma|k \otimes k\> = 0 \ , \ \ \ k=1,2,3.
\end{equation}
To find the remaining spanning vectors consider the following
\begin{equation}  \label{II}
    |\alpha\> = |\beta\> = |1\> + \xi (|2\> + |3\> ) , \ \ \  \xi = r e^{i \phi} \in \mathbb{C} .
\end{equation}
This form is suggested by a symmetric structure of the diagonal blocks of $W_\Gamma$.  One finds
\begin{equation}
    \< \alpha \otimes \beta|W_\Gamma| \alpha \otimes \beta \> = 8 r^2 (r - \cos\phi)^2 ,
\end{equation}
and hence we get a family of vectors $\alpha \otimes \beta$ corresponding to $r = \cos\phi$. There are three natural choices for $\phi = \{0,\frac \pi 4,\frac \pi 2\}$ giving rise to $r=\{1,\frac{1}{\sqrt{2}},0\}$.
If $r=0$, then $|\alpha\> = |1\>$ and it reproduces already known pair $|1 \otimes 1\>$ (cf. Eq. (\ref{I})). Let us take $\phi = \frac \pi 4$ leading to $\xi = \frac{1}{\sqrt{2}}e^{i\pi/4}$.
Consider now the following  modification of (\ref{II})
\begin{equation}  \label{III}
    |\alpha\> =  |1\> + \xi (e^{i \mu} |2\> + e^{-i \mu}|3\> ) , \ \ \  |\beta\> =  |1\> + \xi (e^{-i \mu} |2\> + e^{i \mu}|3\> ) ,
\end{equation}
with $\mu \in \mathbb{R}$, i.e. we introduced a relative phase between $|2\>$ and $|3\>$. One finds
\begin{equation}
    \< \alpha \otimes \beta|W_\Gamma| \alpha \otimes \beta \> = 4[1- \cos(3\mu)] ,
\end{equation}
and hence $ \< \alpha \otimes \beta|W_\Gamma| \alpha \otimes \beta \>=0$ if
\begin{equation}
    \mu = \frac{2\pi n}{3} \ , \ \ n=0,\pm 1,\pm 2,\ldots .
\end{equation}
One finds, therefore,
\begin{equation}  \label{IIa}
    |\alpha\> =  |1\> + \xi (\omega |2\> + \omega^*|3\> ) , \ \ \  |\beta\> =  |1\> + \xi (\omega^*|2\> + \omega|3\> ) ,
\end{equation}
with $\omega = e^{2\pi i/3}$. This construction gives rise to the first three pairs
\begin{equation}
\begin{array}{ll} \vspace{.1cm}
     |\alpha_1\> = |1\> \ , \ & |\beta_1\> = |1\>  \ ,\\ \vspace{.1cm}
 |\alpha_2\> = |1\> + \xi (|2\> + |3\> )  \ , \ & |\beta_2\> = |1\> + \xi (|2\> + |3\> )  \ ,\\ \vspace{.1cm}
    |\alpha_3\> = |1\> + \xi ( \omega |2\> + \omega^* |3\> )  \ , \ & |\beta_3\> = |1\> + \xi (\omega^* |2\> + \omega |3\> )  \ ,
\end{array}
\end{equation}
from the list (\ref{VECTORS}). Simple permutation of $\{|1\>,|2\>,|3\>\}$ provides the remaining pairs from (\ref{VECTORS}).
To check the linear independence of these nine vectors let us construct the $9 \times 9$ matrix $R_1$ out of coordinates of $|\alpha_i \otimes \beta_i\>$, i.e. coordinates of this vector w.r.t. to the product basis $|k \otimes \ell\>$ define the $i$th row of $R_1$:
\begin{equation}
   R_1 =  \left(
\begin{array}{ccccccccc}
 1 & 0 & 0 & 0 & 0 & 0 & 0 & 0 & 0 \\
 0 & 0 & 0 & 0 & 1 & 0 & 0 & 0 & 0 \\
 0 & 0 & 0 & 0 & 0 & 0 & 0 & 0 & 1 \\ \vspace{.1cm}
 1 & \frac{e^{\frac{i \pi }{4}}}{\sqrt{2}} & \frac{e^{\frac{i \pi }{4}}}{\sqrt{2}} & \frac{e^{\frac{i \pi }{4}}}{\sqrt{2}} & \frac{i}{2} & \frac{i}{2} & \frac{e^{\frac{i \pi }{4}}}{\sqrt{2}} & \frac{i}{2} & \frac{i}{2} \\ \vspace{.1cm}
 1 & \frac{e^{-\frac{1}{12} (5 i \pi )}}{\sqrt{2}} & \frac{e^{\frac{11 i \pi }{12}}}{\sqrt{2}} & \frac{e^{\frac{11 i \pi }{12}}}{\sqrt{2}} & \frac{i}{2} & \frac{1}{2} e^{-\frac{1}{6} (i \pi )} & \frac{e^{-\frac{1}{12} (5 i \pi )}}{\sqrt{2}} & \frac{1}{2} e^{-\frac{1}{6} (5 i \pi )} & \frac{i}{2} \\ \vspace{.1cm}
 \frac{i}{2} & \frac{e^{\frac{i \pi }{4}}}{\sqrt{2}} & \frac{i}{2} & \frac{e^{\frac{i \pi }{4}}}{\sqrt{2}} & 1 & \frac{e^{\frac{i \pi }{4}}}{\sqrt{2}} & \frac{i}{2} & \frac{e^{\frac{i \pi }{4}}}{\sqrt{2}} & \frac{i}{2} \\  \vspace{.1cm}
 \frac{i}{2} & \frac{e^{\frac{11 i \pi }{12}}}{\sqrt{2}} & \frac{1}{2} e^{-\frac{1}{6} (i \pi )} & \frac{e^{-\frac{1}{12} (5 i \pi )}}{\sqrt{2}} & 1 & \frac{e^{\frac{11 i \pi }{12}}}{\sqrt{2}} & \frac{1}{2} e^{-\frac{1}{6} (5 i \pi )} & \frac{e^{-\frac{1}{12} (5 i \pi )}}{\sqrt{2}} & \frac{i}{2} \\ \vspace{.1cm}
 \frac{i}{2} & \frac{1}{2} e^{-\frac{1}{6} (i \pi )} & \frac{e^{\frac{11 i \pi }{12}}}{\sqrt{2}} & \frac{1}{2} e^{-\frac{1}{6} (5 i \pi )} & \frac{i}{2} & \frac{e^{-\frac{1}{12} (5 i \pi )}}{\sqrt{2}} & \frac{e^{-\frac{1}{12} (5 i \pi )}}{\sqrt{2}} & \frac{e^{\frac{11 i \pi }{12}}}{\sqrt{2}} & 1 \\ \vspace{.1cm}
 \frac{i}{2} & \frac{i}{2} & \frac{e^{\frac{i \pi }{4}}}{\sqrt{2}} & \frac{i}{2} & \frac{i}{2} & \frac{e^{\frac{i \pi }{4}}}{\sqrt{2}} & \frac{e^{\frac{i \pi }{4}}}{\sqrt{2}} & \frac{e^{\frac{i \pi }{4}}}{\sqrt{2}} & 1 \\
\end{array}
\right).
\end{equation}
One computes
\begin{equation}
    {\rm det} R_1 = \frac{3\sqrt{3}}{16} \left(3 + \frac{5i}{4} \right) \neq 0 ,
\end{equation}
which proves that a set $\{ |\alpha_k \otimes \beta_k \> \}$ spans $\mathbb{C}^3 \otimes \mathbb{C}^3$. Similarly one constructs $R_2$ out of $|\alpha_k \otimes \beta_k^*\>$:
\begin{equation}
    R_2 =
    \left(
\begin{array}{ccccccccc}
 1 & 0 & 0 & 0 & 0 & 0 & 0 & 0 & 0 \\
 0 & 0 & 0 & 0 & 1 & 0 & 0 & 0 & 0 \\
 0 & 0 & 0 & 0 & 0 & 0 & 0 & 0 & 1 \\ \vspace{.1cm}
 1 & \frac{e^{-\frac{1}{4} (i \pi )}}{\sqrt{2}} & \frac{e^{-\frac{1}{4} (i \pi )}}{\sqrt{2}} & \frac{e^{\frac{i \pi }{4}}}{\sqrt{2}} & \frac{1}{2} & \frac{1}{2} & \frac{e^{\frac{i \pi }{4}}}{\sqrt{2}} & \frac{1}{2} & \frac{1}{2} \\ \vspace{.1cm}
 1 & \frac{e^{\frac{5 i \pi }{12}}}{\sqrt{2}} & \frac{e^{-\frac{1}{12} (11 i \pi )}}{\sqrt{2}} & \frac{e^{\frac{11 i \pi }{12}}}{\sqrt{2}} & \frac{1}{2} e^{-\frac{1}{3} (2 i \pi )} & \frac{1}{2} & \frac{e^{-\frac{1}{12} (5 i \pi )}}{\sqrt{2}} & \frac{1}{2} & \frac{1}{2} e^{\frac{2 i \pi }{3}} \\ \vspace{.1cm}
 \frac{1}{2} & \frac{e^{\frac{i \pi }{4}}}{\sqrt{2}} & \frac{1}{2} & \frac{e^{-\frac{1}{4} (i \pi )}}{\sqrt{2}} & 1 & \frac{e^{-\frac{1}{4} (i \pi )}}{\sqrt{2}} & \frac{1}{2} & \frac{e^{\frac{i \pi }{4}}}{\sqrt{2}} & \frac{1}{2} \\ \vspace{.1cm}
 \frac{1}{2} e^{-\frac{1}{3} (2 i \pi )} & \frac{e^{\frac{11 i \pi }{12}}}{\sqrt{2}} & \frac{1}{2} & \frac{e^{\frac{5 i \pi }{12}}}{\sqrt{2}} & 1 & \frac{e^{-\frac{1}{12} (11 i \pi )}}{\sqrt{2}} & \frac{1}{2} & \frac{e^{-\frac{1}{12} (5 i \pi )}}{\sqrt{2}} & \frac{1}{2} e^{\frac{2 i \pi }{3}} \\ \vspace{.1cm}
 \frac{1}{2} e^{-\frac{1}{3} (2 i \pi )} & \frac{1}{2} & \frac{e^{\frac{11 i \pi }{12}}}{\sqrt{2}} & \frac{1}{2} & \frac{1}{2} e^{\frac{2 i \pi }{3}} & \frac{e^{-\frac{1}{12} (5 i \pi )}}{\sqrt{2}} & \frac{e^{\frac{5 i \pi }{12}}}{\sqrt{2}} & \frac{e^{-\frac{1}{12} (11 i \pi )}}{\sqrt{2}} & 1 \\ \vspace{.1cm}
 \frac{1}{2} & \frac{1}{2} & \frac{e^{\frac{i \pi }{4}}}{\sqrt{2}} & \frac{1}{2} & \frac{1}{2} & \frac{e^{\frac{i \pi }{4}}}{\sqrt{2}} & \frac{e^{-\frac{1}{4} (i \pi )}}{\sqrt{2}} & \frac{e^{-\frac{1}{4} (i \pi )}}{\sqrt{2}} & 1 \\
\end{array}
\right) ,
\end{equation}
and computes
\begin{equation}
    {\rm det} R_2 = - \frac{27 \sqrt{3}}{8} \neq 0 ,
\end{equation}
which proves that a set $\{ |\alpha_k \otimes \beta_k^* \> \}$ spans $\mathbb{C}^3 \otimes \mathbb{C}^3$.

\section{Discussion and Generalization to Higher Dimensions}   \label{SEC-C}

Interestingly, EWs and states considered in this paper belong to the well-known class of Bell diagonal operators \cite{magic1,magic2,DC2014,magic3,MagicSimplex1,MagicSimplex2,hiesmayr3}. Introducing a set of unitary Weyl operators in $\mathbb{C}^d$
\begin{equation}
    W_{k\ell} = \sum_{j=0}^{d-1} \omega_d^{jk} |j\>\<j+\ell| , \ \ \ k,\ell=0,1,\ldots, d-1 ,
\end{equation}
with $\omega_d = e^{2\pi i/d}$, one defines a set of generalized Bell states in $\mathbb{C}^d \otimes \mathbb{C}^d$
\begin{equation}
    |\Omega_{k\ell} \> = \oper_d \otimes W_{k\ell} |\Omega_{00}\> ,
\end{equation}
where $|\Omega_{00}\> = \frac{1}{\sqrt{d}} \sum_j |j \otimes j\>$ stands for the canonical maximally entangled state. Now, a Hermitian operator $X$ is a Bell diagonal operator if
\begin{equation}
    X = \sum_{k,\ell=0}^{d-1} x_{k\ell} P_{k\ell} ,
\end{equation}
where $x_{k\ell} \in \mathbb{R}$, and $P_{k\ell} = |\Omega_{k\ell}\>\<\Omega_{k\ell}|$.

\begin{figure}
    \centering
(a)\includegraphics[width=0.47\textwidth]{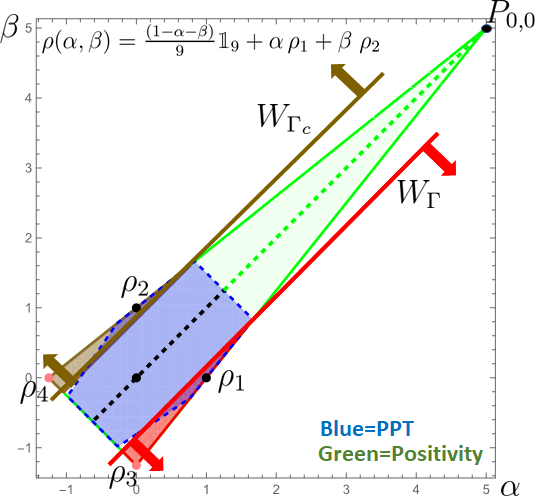}    (b)\includegraphics[width=0.47\textwidth]{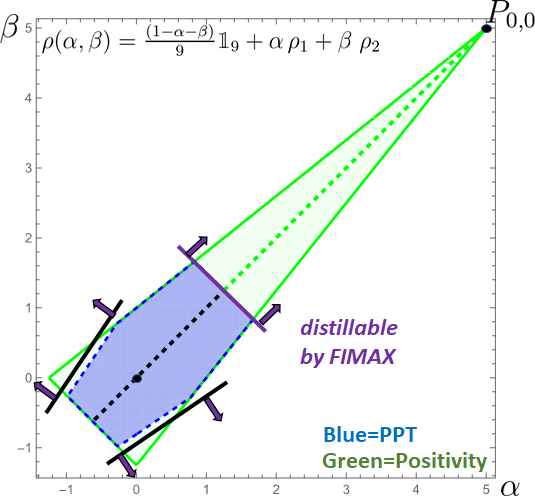}
    \caption{These graphics show a particular slice through the magic simplex~\cite{magic1,magic2,magic3} for dimension $d=3$, which covers the geometry of all Bell diagonal states, i.e. $\sum c_{k,l} P_{k,l}$ with
    $c_{k,l}\geq 0$, $\sum_{k,l} c_{k,l}=1$ and $P_{k,l}=W_{k,l}\otimes \mathbbm{1}_d\; P_{0,0}\; W_{k,l}^\dagger\otimes \mathbbm{1}_d $ with $P_{0,0}$ being any maximally entangled state and $W_{k,l}$ the unitary Wely-operators (``magic'' property). The colored triangle contains all values of $\alpha,\beta$ for which
    $\rho(\alpha,\beta)=\frac{1-\alpha-\beta}{9} \mathbbm{1}_9+\alpha\; \rho_1 + \beta\; \rho_2$ represents a state. Here, $\rho_{\alpha,\alpha}$ equals the isotropic state, for which everything concerning the entanglement properties are known (dashed line in the middle: black $\equiv$ separable state; green $\equiv$ (NPT) entangled state). Due to geometry, we know that there are exactly $9$ equivalent slices in the whole magic simplex (for each of the nine Bell states). (a) The optimal pair of witnesses $W_\Gamma/W_{\Gamma_s}$ are in this picture ``parallel'' lines (red/brown). The region $(\alpha,\beta)$ for which they detect entanglement is colored red or brown, respectively. Those witnesses also detect states which are PPT (blue area). The states $\rho_{3/4}$ are those of Eq.~(\ref{CESstates}), which span the completely entangled subspace (CES) of $\mathbb{C}^3\otimes\mathbb{C}^3$. (b) Shows the region for which the distillation protocol  FIMAX~\cite{DistillationProtocol}, which is the currently best-performing protocol for Bell-diagonal states~\cite{DistillationProtocolPerformance}, is able to distill in a one-shot, two-copy scenario NPT-entangled states. Notable, the region of NPT entangled states for which the protocol is not successful is not ``parallel'' to the pair of optimal entanglement witnesses, but to the border of PPT.}
    \label{Fig1}
\end{figure}

One finds for a mirrored pair: $W_\Gamma =  \sum_{k,\ell=0}^{d-1} w_{k\ell} P_{k\ell} $ and
$W_{\Gamma_c}=\sum_{k,\ell=0}^{d-1} \tilde{w}_{k\ell} P_{k\ell}$, with
\begin{equation}
    w_{k\ell} =  \left( \begin{array}{ccc} 2 & -1 & -1 \\ -1 & 5 & 5 \\ -1 & 5 & 5 \end{array} \right) \ ,\ \ \ \   \tilde{w}_{k\ell} =  \left( \begin{array}{ccc} 2 & 5 & 5 \\ 5 & -1 & -1\\ 5 & -1 & -1 \end{array} \right) .
\end{equation}
Similarly one finds $\rho_1 =  \sum_{k,\ell=0}^{d-1} c_{k\ell} P_{k\ell} $ and
$\rho_2 =\sum_{k,\ell=0}^{d-1} \tilde{c}_{k\ell} P_{k\ell}$, with
\begin{equation}
    c_{k\ell} = \frac 15 \left( \begin{array}{ccc} 1 & 1 & 1 \\ 1 & 0 & 0 \\ 1 & 0 & 0 \end{array} \right) \ ,\ \ \ \   \tilde{c}_{k\ell} = \frac 15 \left( \begin{array}{ccc} 1 & 0 & 0 \\ 0 & 1 & 1\\ 0 & 1 & 1 \end{array} \right) .
\end{equation}
It is therefore clear that $W_\Gamma$ gives rise to a CES spanned by $\{|\Omega_{01}\>,|\Omega_{02}\>,|\Omega_{10}\>,|\Omega_{12}\>\}$. Similarly, its mirrored partner
gives rise to a CES spanned by $\{|\Omega_{11}\>,|\Omega_{12}\>,|\Omega_{21}\>,|\Omega_{22}\>\}$. Interestingly, states supported on these CESs
\begin{equation}\label{CESstates}
    \rho_3 = \frac 14 (P_{01}+P_{02}+P_{10}+P_{20}) \ , \ \ \ \rho_4 = \frac 14 (P_{11}+P_{12}+P_{21}+P_{22}) \ ,
\end{equation}
satisfy
\begin{equation}
    {\rm Tr}(W_{\Gamma} \rho_3) = {\rm Tr}(W_{\Gamma_c} \rho_4) = -1 ,
\end{equation}
however, they are not PPT, cf. the Figure \ref{Fig1}~(a). { Recently, a novel distillation protocol, FIMAX~\cite{DistillationProtocol}, is introduced which surpasses all other existing protocols for Bell diagonal states~\cite{DistillationProtocolPerformance}. The results for this family of states are displayed in Figure \ref{Fig1}~(b). Both states, $\rho_3,\rho_4$, can be distilled by local operations and classical communication (LOCC) to a maximally entangled state, e.g. $P_{0,0}$. Only, the highly mixed states close to the PPT border cannot be distilled by the one-shot, two-copy protocol FIMAX.}

The presented structure of mirrored EWs in $3 \otimes 3$ suggests generalization for $d \otimes d$ with $d>3$.
Let us consider the state space given by the magic simplex~\cite{magic1,magic2,magic3}, i.e.
\begin{equation}
\mathcal{W}_d=\left\lbrace \rho=\sum_{k,\ell=0}^{d-1} c_{k\ell}\; P_{k\ell}\;|\;c_{k\ell}\geq 0,\sum_{k,\ell=0}^{d-1} c_{k\ell}=1\right\rbrace \;.
\end{equation}
Obviously, a complete orthonormal basis of Bell states can be also achieved by arbitrary unitary operators, however, the Weyl relations imply strong entanglement features. In Ref.~\cite{hiesmayr4}, this was studied in detail and showed that the Weyl-relations lead to the maximal volume of PPT-entangled states within any simplex. A similar behavior is also observed when considering the performance of distillation protocols on magic simplex states versus non-magic simplex states Ref.~\cite{DistillationProtocolPerformance}.   Therefore, we present here only the results with respect to the magic simplex.

In this case, it has been proven~\cite{magic2} that any EW can be brought into the Bell diagonal form, i.e. $W=\sum \kappa_{k\ell} P_{k\ell}$. It is well known that the geometry given by the Weyl operators allows for $d+1$ lines, namely a set of Bell states form a line when they are connected by a single Weyl operator. Moreover, an equal mixture of those Bell states forming a line (in the phase-space), for example, $\frac{1}{d}\sum_{l=0}^{d-1} P_{0,l}$, and is always separable. Those states form the outmost separable states in the magic simplex and define a so-called kernel, for which any state inside has to be separable. For even dimensions, there are also substructures therefor let us here concentrate on odd dimensions. Furthermore, any set of states $\{c_{k,l}| \;\;\textrm{all}\quad c_{k,l}\in[0, \frac{1}{d}]\}$ forms a so-called enclosure polytope for which any state outside is certainly not PPT, i.e. entangled. So the region between the kernel and the enclosure polytope is the region for searching for bound entangled states. For $d=3$ and $d=4$ the whole region of bound entangled states within the magic simplex have been effectively detected~\cite{MagicSimplex2,hiesmayr3}.

\begin{figure}
    \centering
\includegraphics[width=0.98\textwidth]{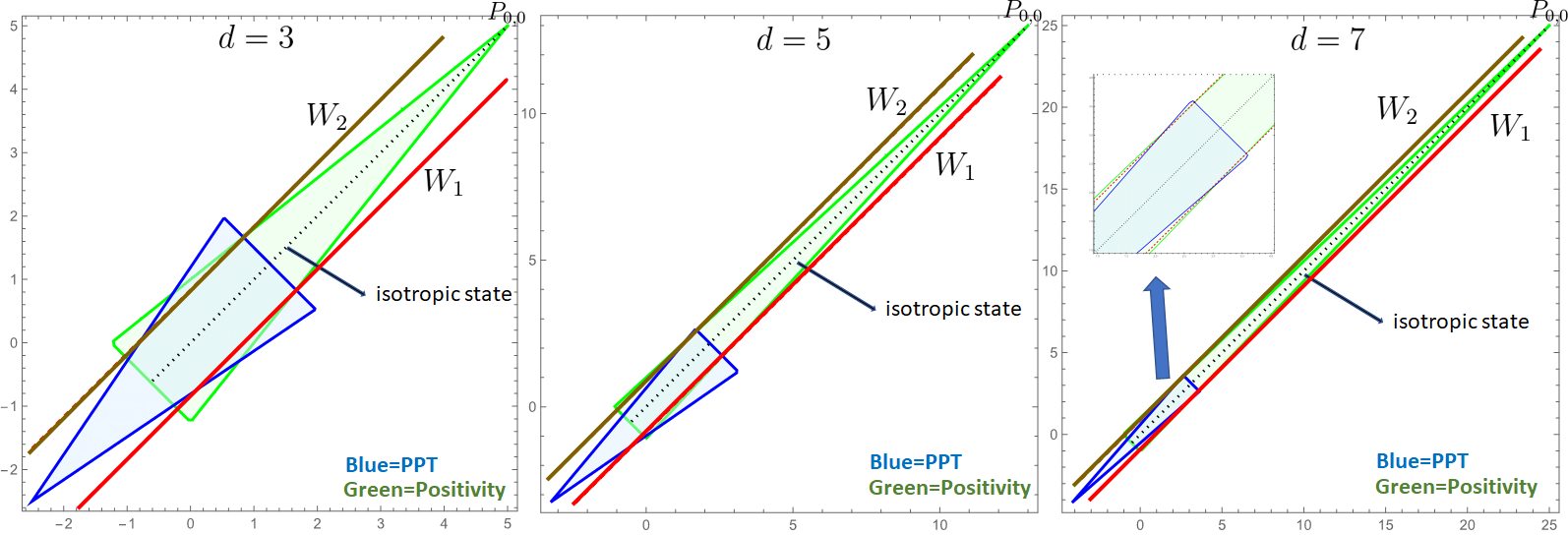}
    \caption{These graphics compare one slice through the magic simplex~\cite{magic1,magic2,magic3} for different dimension $d=3,5,7$. The mirrored witness pairs $\{W_1,W_2\}$ (red/gold lines) are always ``parallel'' to the isotropic state $\rho_{iso}(p)=p \,P_{0,0}+(1-p)\, \frac{1}{d^2} \mathbbm{1}_{d^2}$ (black dotted line). In all dimensions, both witnesses detect bound entangled states. In the case of $d=3,7$, the PPT states (blue and green) are symmetric around the isotropic state, and by that, they detect the same amount of bound entangled states. However, for $d=5$ the PPT region (blue) is not symmetric with respect to the positivity region (green); both EWs detect bound entangled states, but not the same amount. Thus symmetry on the Hermitian but not positive space is not in one-to-one correspondence with symmetry with respect to transposition in one subsystem and positivity.}
    \label{figgeometrycomparison}
\end{figure}

In dimension $d=3$ states $\rho_1$ and $\rho_2$ where both PPT and as depicted in Fig.~\ref{figgeometrycomparison} the positivity (green) and PPT region (blue) are symmetric to each other. In dimension $d=5$ this is not the case, however, for $d=7$ we recover this symmetry again (see Fig.~\ref{figgeometrycomparison}).
This shows a symmetry that exists on mirrored EWs but lacks in the state space, which does not necessarily exhibit the property with respect to the partial transposition.

Let us single out one Bell state, e.g. $P_{0,0}$, for which we will choose $\kappa_{k,l}>0$. Then we have $d^2-1$ remaining $\kappa_{k,l}$ for which maximally $(d-1)(d-1)$ can be negative. In dimension $d=3$, we found $4$, which equals half of the number of the remaining weights. Taking this rule, we see that it does not match in higher dimensions. For dimension $d=5$, we can only have $12$ negative values to keep the symmetry between the paired witnesses, e.g.
 \begin{eqnarray}
 &&W_1\to \kappa_{k\ell} = \left(\begin{array}{ccccc}
 4&-1&-1&-1&-1\\
 -1&-1&9&9&9\\
 -1&9&-1&9&9\\
 -1&9&9&-1&9\\
 -1&9&9&9&-1\\
 \end{array}\right)\;,\quad W_2\to \kappa_{k\ell} = \left(\begin{array}{ccccc}
 4&9&9&9&9\\
 9&9&-1&-1&-1\\
 9&-1&9&-1&-1\\
 9&-1&-1&9&-1\\
 9&-1&-1&-1&9
 \\
 \end{array}\right)\nonumber\\
 &&
 W_3\to \kappa_{k\ell} = \left(\begin{array}{ccccc}
 4&-1&-1&-1&-1\\
 9&9&-1&9&-1\\
 9&9&9&-1&-1\\
 9&-1&-1&9&9\\
 9&-1&9&-1&9\\
 \end{array}\right)\;,\quad W_4\to \kappa_{k\ell} = \left(\begin{array}{ccccc}
 4&9&9&9&9\\
 -1&-1&9&-1&9\\
 -1&-1&-1&9&9\\
 -1&9&9&-1&-1\\
 -1&9&-1&9&-1\\
 \end{array}\right)\;,
 \end{eqnarray}
 where $\{W_1,W_2\}$ and $\{W_3,W_4\}$ form each a pair of mirrored witnesses

\begin{equation}
    W_1 + W_2 = 8\; \oper_5 \otimes \oper_5 \ , \ \ \ W_3 + W_4 = 8\; \oper_5 \otimes \oper_5 \ .
\end{equation}

Given this line structure, it is easy to construct states detected by those EWs, i.e.
 \begin{eqnarray}
 &&\rho_1\to c_{k\ell} =  \frac{1}{13}\left(\begin{array}{ccccc}
 1&1&1&1&1\\
 1&1&0&0&0\\
 1&0&1&0&0\\
 1&0&0&1&0\\
 1&0&0&0&1\\
 \end{array}\right)\;,\quad \rho_2\to c_{k\ell} =  \frac{1}{13}\left(\begin{array}{ccccc}
 1&0&0&0&0\\
 0&0&1&1&1\\
 0&1&0&1&1\\
 0&1&1&0&1\\
 0&1&1&1&0
 \\
 \end{array}\right)\nonumber\\
 &&
 \rho_3\to c_{k\ell} =  \frac{1}{13}\left(\begin{array}{ccccc}
 1&0&0&0&0\\
 0&0&1&0&1\\
 0&0&0&1&1\\
 0&1&1&0&0\\
 0&1&0&1&0\\
 \end{array}\right)\;,\quad \rho_4\to c_{k\ell} =  \frac{1}{13}\left(\begin{array}{ccccc}
 1&0&0&0&0\\
 1&1&0&1&0\\
 1&1&1&0&0\\
 1&0&0&1&1\\
 1&0&1&0&1\\
 \end{array}\right)\;.
 \end{eqnarray}
Indeed, one hinds ${\rm Tr} W_i \rho_i= - \frac{8}{13}$.

%\frac{\mu}{(d+1)2+1}=- [TO BE CHECKED IF THIS IS A GENERAL FORMULA].

\begin{figure}
    \centering
    (a)\includegraphics[width=0.47\textwidth]{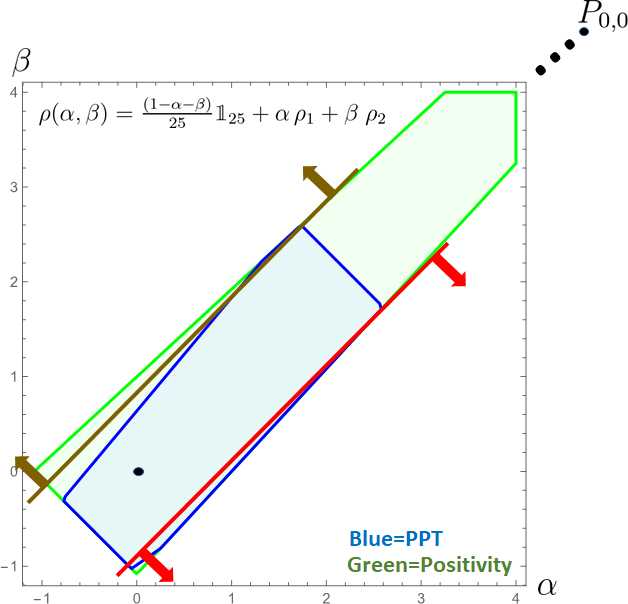}
    (b)\includegraphics[width=0.47\textwidth]{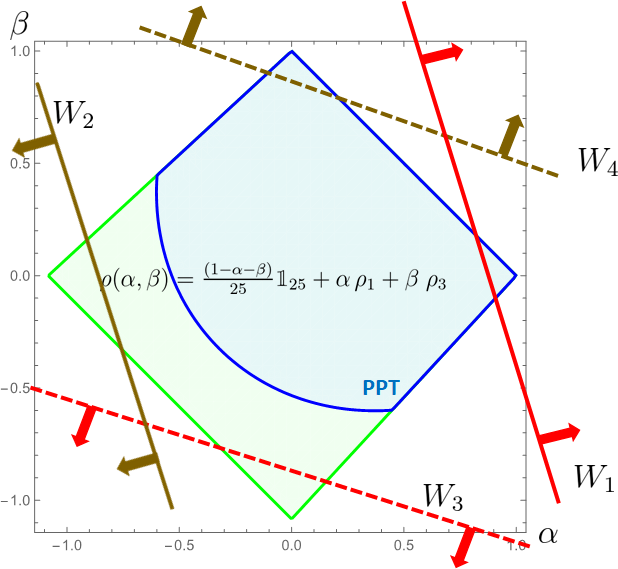}
    \caption{These graphics show two different slices through the magic simplex~\cite{magic1,magic2,magic3} for dimension $d=5$. The (blue) area covers all PPT states. The state space is symmetric in the parametrization $\alpha,\beta$, however, the PPT property is only in (b) symmetric. The pair of EWs (red/brown) are depicted, and for (a) we find that both detect bound entangled states, whereas in (b) only one detects bound entangled states.}
    \label{figdimfive1}
\end{figure}

In $d=5$, the states $\rho_1$ and $\rho_3$ are PPT, but not $\rho_2$ and $\rho_4$, thus the mirrored property has no one-to-one relation, though the states and EWs share the same ``geometry''. Still both pairs are non-decomposable since they generally do detect bound entangled states, which is displaced in Fig.~\ref{figdimfive1}(a) by considering the state family $\rho(\alpha,\beta)=\frac{1-\alpha-\beta}{25} \mathbbm{1}_{25}+\alpha\, \rho_{1}+\beta\, \rho_{2}\equiv \frac{1-\alpha-\beta}{25} \mathbbm{1}_{25}+\alpha\, \rho_{4}+\beta\, \rho_{3}$. The PPT-subspace is however not symmetric. We can make the subspace symmetric by considering the state family of the two PPT states, i.e.  $\rho(\alpha,\beta)=\frac{1-\alpha-\beta}{25} \mathbbm{1}_{25}+\alpha\, \rho_{1}+\beta\, \rho_{3}$, which is displayed in Fig.~\ref{figdimfive1}(b). Here, only one of the EW of a pair is detecting bound entanglement, see Fig.~\ref{figdimfive1}(a).

\section{Summary and Outlook}  \label{SEC-SUM}

There are two standard methods to address the problem of separability. Either directly on the state space, i.e. on bounded positive operators, or with the help of entanglement witnesses, Hermitian non-positive operators. The NP-hardness of the separability problem stems from states which are PPT (positive under partial transposition). On the other hand witnesses that are decomposable cannot detect those PPT-entangled states. In this paper, we provided a counterexample to the conjecture that for a pair of optimal mirrored entanglement witnesses, only one in each pair can be non-decomposable. Interestingly both EWs from a mirrored pair $(W_\Gamma,W_{\Gamma_c})$ in $3 \otimes 3$ are non-decomposable optimal (they enjoy bi-spanning property \cite{Lew}). The construction from $d=3$ is then generalized to mirrored EWs in $d=5$ and $d=7$. Again both witnesses from each mirrored pair turned out to be non-decomposable. Surprisingly, analyzing the geometry of space of states, we
find that some cross-sections containing maximally mixed state and maximally entangled state display intriguing asymmetry, that is, for  $d=3$ and $d=7$, a subset of PPT states is perfectly symmetric w.r.t. the line connecting maximally mixed and maximally entangled states. However, the symmetry is lost for $d=5$.
%an asymmetry  between the set of PPT  states and positivity in $d=5$ but not for $d=3$ and $d=7$.
{This makes it clear that we do not yet have the right relationship
between the properties of the entanglement witnesses and the properties of the state space.}

There are still several interesting questions to address. Are entanglement witnesses $W_i$ ($i=1,2,3,4$) optimal (or even non-decomposable optimal)? To prove optimality (or non-decomposable optimality), it would be sufficient to find $25$ product vectors with spanning (bi-spanning) property. It is rather straightforward to find $15$ product vectors $|\alpha \otimes\beta\>$ such that $|\alpha\>$ and $|\beta\>$ belong to a set of $30$ vectors from six MUBs. However, it is not at all obvious how to complete this set to a complete set of $25$ span vectors.

An important question is how to generalize the above construction for higher dimensions? In Ref.~\cite{jpa22}, we analyzed the question of how many mutually unbiased bases are needed to detect bound entangled states. It was shown that if the number of MUBs (measurements) is greater than $d/2+1$,  one can provide a construction of a suitable non-decomposable EW. Could we do better? That is, could we provide a construction with a smaller number of MUBs? We plan to address the above problems in future work.

 \vspace{1cm}

\section*{Acknowledgements}
DC was supported by the Polish National Science Centre project No. 2018/30/A/ST2/00837.   JB was supported by National Research Foundation of Korea (NRF-2021R1A2C2006309, NRF-2020K2A9A2A15000061), Institute of Information \& communications Technology Planning \& Evaluation (IITP) grant (the ITRC Program/IITP-2021-2018-0-01402).  BCH acknowledges gratefully that this research was funded in whole, or in part, by the  Austrian Science Fund (FWF) project P36102-N (Grant
DOI: 10.55776/P36102). For the purpose of open access, the author has applied a CC BY public copyright licence to any Author Accepted Manuscript version arising from this submission.

\end{document}